\documentclass[conference]{IEEEtran}

\usepackage{enumerate}
\usepackage{amsmath,amssymb,bm,amsfonts}
\usepackage{graphics,graphicx,epsfig,subfigure}
\usepackage{algorithm,algorithmic}
\usepackage{theorem}
\usepackage{threeparttable}
\usepackage{stfloats}
\usepackage{cite}
\usepackage{multirow}
\usepackage{mathrsfs}
\usepackage{amsmath,bm,bbm}
\usepackage{color}
\usepackage{multirow}
\usepackage{subeqnarray}
\usepackage{cases}
\usepackage{colortbl}
\usepackage{wasysym}
\usepackage[dvipsnames]{xcolor}
\usepackage{url}

\newtheorem{mytheorem}{Theorem}

\newtheorem{myproof}{Proof}

\newtheorem{myproperty}{Property}
\begin{document}
\newcommand*{\QEDA}{\hfill\ensuremath{\blacksquare}}
\def\arrow{{\rightarrow}}
\def\N{{\mathcal{N}}}
\def\B{{\mathcal{B}}}
\def\E{{\mathcal{E}}}
\def\I{{\mathcal{I}}}
\def\diag{{\textrm{diag}}}
\def\i {{ -i}}
\def\ci{\perp\!\!\!\perp} 
\newcommand\independent{\protect\mathpalette{\protect\independenT}{\perp}} 
\def\independenT#1#2{\mathrel{\rlap{$#1#2$}\mkern2mu{#1#2}}}

\title{Distributed Frequency Offsets Estimation}

\author{
	{Jian Du and Shaodan Ma }\\
}

\maketitle

\IEEEpeerreviewmaketitle
\begin{abstract}
In this paper, we provide a distributed frequency offset estimation algorithm based on a variant of belief propagation (BP).
Each agent in the network pre-compensates its carrier frequency individually so that there is no frequency offset from the desired carrier frequency between each pair of transceiver.
The pre-compensated offset for each agent is computed in a distributed fashion in order to be adaptive to the distributed network.
The updating procedure of the variant of BP is designed in a broadcasting fashion to reduce communication burden.
It is rigorously proved that the proposed algorithm is convergence guaranteed.
Simulations show that this method achieves almost the optimal frequency compensation accuracy with an error approaching the Cram\'{e}r-Rao lower bound.
\end{abstract}

%
%
\section{Introduction}
In wireless communication networks, frequencies synthesized from
independent oscillators could be different from each other due
to variation of oscillator circuits, and this  difference  is known as carrier frequency offset (CFO).
The received signal impaired
by CFO between transmitter and
receiver leads to a continuous rotation of symbol constellation,
resulting in degradation of system capacity and bit error rate
 \cite{8081821, cai2010cfo, du2008multiple}.
To overcome this problem, traditional CFO estimation and compensation has been studied by centralized processing, i.e., by gathering all the information in a
central processing unit, CFOs are estimated at the receiver
and then fedback to corresponding transmitters to adjust the
offsets.
However, it is known that centralized processing are not scalable to large-scale networks, e.g, in the context of beamforming \cite{4202181}, clock synchronization \cite{clockJ, clockconf}, and power state estimation \cite{pmudu, powerconf}.
Recently, a
belief propagation based fully distributed
CFO estimation and compensation method is proposed in \cite{du2013network}, which only involves
local processing and information exchange between direct
neighbors.
Although this
 method converges fast to the centralized optimal solution,
the number of messages involved  at each iteration  grows quadratically as the number of agents (transmitters and receivers) in network increases, leading to information network congestion.
To overcome this problem, in this paper, we take a step further and propose a novel distributed algorithm named as linear scaling belief propagation (LSBP) for its linear scalability to network density.
We apply LSBP to network-wide CFO estimation for communication networks with arbitrary topologies. It is shown that the total number of messages at each iteration simply equals to the number of agents.

From the theoretical analysis perspective, the convergence properties are analyzed for LSBP.
Note that though BP has gained great success in many applications,  it is found that BP may diverge if the network topology contains circles,
and the necessary and sufficient convergence condition is still an open problem.
In contrast, the analytical analysis of the proposed LSBP algorithm shows that LSBP is convergence guaranteed for arbitrary  network topology.
Besides, even with different initial values, the LSBP converges to a unique point.
The above theoretical analysis is also verified by simulations, and it is shown that the proposed LSBP algorithm converges quickly with the estimation mean-square-error (MSE) approaching the Cram\'{e}r-Rao lower bound (CRLB).



\section{Problem Formulation and Modeling
}\label{model}

The communication network is represented
by an undirected graph
$\mathcal{G}=(\mathcal{V},\mathcal{E})$,
where $\mathcal{V}=\{1,\ldots, N\}$ is the set of agents, and $\mathcal{E}\subseteq \mathcal{V}\times \mathcal{V}$ is the set of communication links between agents.
Neighbors of agent $i$
are denoted by $\mathcal{B}(i)\triangleq \{j\in \mathcal{V}| (i,j)\in \mathcal{E}\}$.
Let $f_{i,j}$ be the CFO between $i$ and $j$, then the pre-compensated frequency shift at agent $i$ and at agent $j$, i.e., $f_{i}$ and $f_{j}$, should satisfy
$f_i + f_j = f_{i,j}$.
In practice, we can only obtain the measurement or estimate \cite{DopplorSpeed,JWChen} of $f_{i,j}$, denoted as $r_{i,j}$, between neighboring agents $\{i,j\}\in \E$.
Thus, we have
\begin{equation} \label{linearSPEED}
r_{i,j}
=  f_i +  f_j + n_{i,j},
\end{equation}
where $n_{i,j}$ is the estimation error.
{It is known that  the maximum
likelihood  estimates of $n_{i,j}$ is asymptotically Gaussian distributed \cite{JWChen}, that is,
$n_{i,j}\sim \mathcal{N}( n_{i,j};  0, \sigma^2_{i,j})$.}

Implementing centralized estimator, not only requires bringing all  $r_{i,j}$ and $\sigma_{i,j}^2$ to a central computing unit, but also needs the  topology of $\mathcal{G}$.
Thus, the  centralized estimator is not scalable with network size, which causes heavy communication  burden by transmitting data from network border to control unit.
Therefore, distributed estimation, where each agent performs estimation with local information, sounds promising.

In the following,  BP  is introduced first for estimation of pre-compensated frequency shift. Inspired by BP, a distributed estimation algorithm named as linear scaling BP (LSBP), which has low communication overhead and is convergence guaranteed, is then analyzed.
{Notice that communication scheme \cite{robust} that is robust to carrier frequency offset can be adopted for message exchange before frequency offset are compensated \cite{networkCFOconf}.}
\subsection{Belief Propagation Algorithm
}
With BP \cite{du2016convergence} algorithm for linear Gaussian model, at every iteration, each agent sends a (different) message
to each of its neighbors and receives a message from each
neighbor.
The message from agent $j$ to agent $i$ is defined as
the product of the local function $\N(r_{i,j};f_i+f_j,\sigma_{i,j}^2)$ with messages received from all neighbors except $ i$,
and then maximized over all involved variables except $ f_i$.
Mathematically, it is defined as
\begin{equation}  \label{BPf2vs1}
\begin{split}
m^{(l)}_{j \arrow i}(f_i)
= &\mathop{\max}
\limits_{f_{j}}
\quad \N(r_{i,j};f_i+f_j,\sigma_{i,j}^2)\\
& \times \prod_{k\in\B(j)\setminus i} m^{(l-1)}_{k \arrow j}(f_j).
\end{split}
\end{equation}

The message $m^{(l)}_{j \arrow i}(f_i)$ is computed and exchanged among neighbors.
One possible  scheduling for message exchange is  that all  agents perform local computation and message exchange in parallel.
In any round of message exchange, a belief of $ f_i$ can be computed at each agent $i$ locally,  as the product of all the incoming messages from neighbors, which is given by
\begin{equation} \label{BPbelief}
b^{(l)}(f_i)
 =  \prod_{ j\in\B(i)} m^{(l)}_{ j \arrow i}( f_i).
\end{equation}
The belief $b^{(l)}(f_i)$ serves as the approximation of the optimal centralized estimator.
Therefore, the estimate of $ f_i$ in the $l^{th}$ iteration can be computed by
\begin{equation}\label{est1}
\hat {f}_i^{(l)}
 =  \mathop{\max} \limits_{f_i} b^{(l)}(f_i).
\end{equation}

It is apparent that the  outgoing messages, i.e., $ m^{(l)}_{ j \arrow i}( f_i)$, to different neighbors are different, and thus, huge amount of information is broadcasted in the network. Such problem is especially serious in dense traffic and leads to information network traffic congestion \cite{BroadcastinginVanet, LowCongestionControl}.

\subsection{Message Computation for Linear Scaling BP}
To address the above
problems,   we use a variant of BP for distributed frequency offset estimaiton.
It not only guarantees the iterative updating convergence
but also has the property that
the  amount of information exchange among agents
 is linear to the traffic density.
The message updating equation is defined as
\begin{equation}  \label{BBPf2vs1}
\tilde{m}^{(l)}_{j \arrow i}(f_i)
=  \mathop{\max}\limits_{f_j}
\N(r_{i,j};f_i+f_j,\sigma_{i,j}^2)
b^{(l-1)}_{j}(f_j),
\end{equation}
and the outgoing message is
\begin{equation} \label{BBPbelief}
\tilde{b}^{(l)}(f_i)
=  \prod_{ j\in\B(i)} \tilde{m}^{(l)}_{ j \arrow i}( f_i).
\end{equation}
Note that (\ref{BBPf2vs1}) differs from the standard BP of (\ref{BPf2vs1}) in that
each agent broadcasts  $\tilde{b}^{(l-1)}(f_j)$ to all its neighbors at one time, then $\tilde{m}^{(l)}_{j \arrow i}(f_i)$ is computed at node $i$ and then the belief $\tilde{b}^{(l)}(f_i)$ can be obtained according to (\ref{BBPf2vs1}).
Because  the message needs to be transmitted at each iteration equals the number of agents,
the proposed method is named as linear scaling BP (LSBP).
Next, the explicit message expression of LSBP is computed.

To start the recursion,
in the first round of message exchange, the initial incoming message is settled as  $b^{(l-1)}_{j}(f_j)=\mathcal{N}( f_j; \mu^{(0)}_{j},  P_{j}^{(0)})$,
with $ P_{j}^{(0)} >0$ and $\mu^{(0)}_{j}$ can be arbitrary value.
Since  $\N(r_{i,j};f_i+f_j,\sigma_{i,j}^2)$ is a Gaussian pdf,
according to (\ref{BBPf2vs1}),  $\tilde{m}^{(1)}_{j \arrow i}(f_i)$ is still a Gaussian function.
In addition, $\tilde{b}^{(1)}(f_i)$, being the product of Gaussian functions in (\ref{BBPbelief}), is also a Gaussian function \cite{du1}.
Consequently, in LSBP, during each round of message exchange, all the messages are Gaussian functions, and
only the mean  and the variance need to be exchanged between neighbors.

At this point, we can compute the messages of LSBP at any iteration.
In general, in the $l^{\textrm{th}}$ ($l=2,3,\cdots$) round of message exchange, agent $i$ with the available  message
$b^{(l-1) }_{ j } (f_j)\propto
\N(f_j;\mu^{(l-1)}_{j}, P_j^{(l-1)})$ from its neighbors, computes the outgoing messages via (\ref{BBPf2vs1}).
By putting the explicit expression of
$b^{(l-1)}_{ j } (f_j)$  into
(\ref{BBPf2vs1}) and after some tedious but straightforward  computations, we have
$\tilde{m}^{(l)}_{j \arrow i}(f_i)\propto
\N(f_i; \eta^{(l)}_{j\arrow i}, C_{j\arrow i}^{(l)})$
in which
\begin{equation} \label{f2fCBBP}
  C_{j\arrow i}^{(l)}
 =
  \sigma_{i,j}^2 +
P_j^{(l-1)},
\end{equation}
and
\begin{eqnarray}\label{f2fvBBP}
\eta^{(l)}_{j\arrow i}
&=&
r_{i,j}+ \mu^{(l-1)}_{j}.
\end{eqnarray}

Furthermore, during each round of message exchange, each agent  computes the belief for $f_i$
via (\ref{BBPbelief}), which can be easily shown to be
$\tilde{b}_{i}^{(l)}(  f_i)\propto\mathcal{N}( f_i;  \mu_i^{(l)},   P_{i}^{(l)})$, with variance
\begin{equation} \label{beliefPBBP}
 P_{i}^{(l)} =   \big[ \sum_{j\in\B(i)}\big[C_{j\arrow i}^{(l)}\big]^{-1}\big]^{-1} ,
\end{equation}
and mean
\begin{equation}\label{beliefuBBP}
 \mu_i^{(l)} =  P_{i}^{(l)} \big\{
 \sum_{j\in\B(i)}
 \big[ C_{j\arrow i}^{(l)}\big]^{-1} \eta^{(l)}_{j\arrow i}\big\}.
\end{equation}

The updating is iterated between (\ref{f2fCBBP}), (\ref{f2fvBBP}) and
(\ref{beliefPBBP}), (\ref{beliefuBBP}) at each agent in parallel.
One way to terminate the iterative algorithm is that all agents stop updating when a predefined maximum number of iterations $l_{\textrm{max}}$ is reached.
Since LSBP is convergence guaranteed as proved in the next section, the termination can also be implemented once the algorithm converged.
The LSBP algorithm is summarized in Algorithm 2.


It can be easily concluded that in contrast to BP algorithm, with which the amount of messages need to be computed and transmitted by each agent at each iteration is proportional to the number of neighbors, with LSBP each agent only needs to compute and transmit one pair of mean and variance to all its neighbors.
Therefore, LSBP is scalable with traffic density.
Moreover, in a limit case where $\mathcal{G}$ is a fully connected graph, i.e., $|\mathcal{B}(i)|=N-1$,
 the number of messages exchanged in the network with BP  is $(N-1)N$.
Thus the total number
of messages,  grows quadratically when the
agent number $N$ increases, leading to information network
congestion.
While with LSBP, it is only $N$.
Therefore, the number of messages involved in BP increases much faster than that with LSBP which leaves the network vulnerable to information congestion.
To get further insights of the  proposed LSBP algorithm, its convergence property is studied in the following section.

\section{Convergence Analysis for LSBP}\label{analysis}
As BP may diverge if the network topology contains circles \cite{du2,du3}, which is often the case in communication networks, BP is not reliable.
In this section, we analytically proved that the proposed LSBP algorithm is convergence guaranteed
with feasible initial values,  and $\mu^{(l)}_{j}$ and $ P_{j}^{(l)} $ converge to the same fixed point respectively even with different initial value pairs  $\mu^{(0)}_{j}$ and $ P_{j}^{(0)}$.
Due to the estimate by LSBP  shown in (\ref{beliefuBBP}) depends on  $P_{i}^{(l)} $ and
$\eta^{(l)}_{j\arrow i}$, we first prove
the convergence of $ P_{i}^{(l)}$ and then $\eta^{(l)}_{j\arrow i}$.

\subsection {Convergence of Message Variance} By substituting (\ref{f2fCBBP}) into (\ref{beliefPBBP}), the updating equation of $P_{i}^{(l)}$ is given by
\begin{equation} \label{beliefP2}
 \big[ P_{i}^{(l)}\big]^{-1} =   \sum_{j\in\B(i)}\big[  \sigma_{i,j}^2 +
P_j^{(l-1)} \big]^{-1}.
\end{equation}
Let $\bm p^{(l)}$ be  a vector containing of all the message variance at the $l^{\textrm{th}}$ iteration, i.e., $\bm p^{(l)}\triangleq[[P_2^{(l)}]^{-1},[P_3^{(l)}]^{-1},\ldots, [P_N^{(l)}]^{-1}]^T$ and  define an evolution function $ \mathbb{F} $ as $\bm p^{(l+1)}=  \mathbb{F}(\bm p^{(l)}) $.
We will say that a $\bm p^{(0)}>0$ is
a feasible initial value if $\bm p^{(0)}>0$ satisfies
$\mathbb{F}(\bm p^{(0)})\geq \bm p^{(0)})$
or $\mathbb{F}(\bm p^{(0)})\leq \bm p^{(0)}$.
Notice that one easy obtained feasible
$\bm p^{(0)}$ is by setting $\big[ P_{i}^{(l)}\big]^{-1}=0$.
Though the  expression for the message covariance of LSBP is different from that of BP, with the same methodology for the analysis of BP, we can
show that the function $\mathbb{F}(\cdot)$ has the following properties
for arbitrary $\bm p^{(0)}>0$.
The detail proof is omitted due to space limitation and interested readers may refer to the analysis of BP in \cite{du2016convergence, du3} for the proof.
\begin{myproperty} \label{P-Cov}
The following claims hold with $l\in \{0,1,\cdots\} $:\\
P\ref{P-Cov}-1. Positive limited range: $\mathbb{F}(\bm 0 ) >\mathbb{F}(\bm p^{(l)} )>0$.\\
P\ref{P-Cov}-2. Scalability: $\forall \alpha>1, \alpha \mathbb{F}(\bm p^{(l)})  > \mathbb{F}(\alpha \bm p^{(l)} )$.\\
P\ref{P-Cov}-3. Monotonicity: if $\bm p^{(l)}  \geq \tilde{\bm p}^{(l)}$ then
$\mathbb{F}(\bm p^{(l)} ) \geq \mathbb{F}(\tilde{\bm p}^{(l)})$.
\end{myproperty}

For arbitrary iterative function with the above properties, it is shown recently in \cite{du2016convergence, du3} that the iterative function converges to a fixed point at a superlinear convergence rate. Thus, we have the following theorem.  .
\begin{mytheorem}\label{ConConvTh}
With arbitrary feasible initial value $P_i^{(0)}>0$, the belief variance $ P_i^{(l)}$ of LSBP shown in (\ref{beliefP2})  converges to a unique
fixed point at a superlinear rate for a specific network topology.
\end{mytheorem}

Next, we focus on the convergence property of the estimate $\mu_i^{(l)}$ with the conclusion that $P_i^{(l)}$ has converged.
\subsection {Convergence of Message Mean}
Suppose the converged value of
$P_j^{(l)}$ is $P_j^{\ast}$, then following (\ref{f2fCBBP}), we have
$C_{j\arrow i}^{(l)}=\sigma_{i,j}^2 +
P_j^{\ast}$.
Thus, $C_{j\arrow i}^{(l)}$ is also convergence guaranteed, and then the converged value is denoted by
$C_{j\arrow i}^{\ast}$.
Putting $C_{j\arrow i}^{\ast}$ into (\ref{beliefPBBP}) and substituting the result into (\ref{beliefuBBP}), we have
\begin{equation}\label{dsbp}
\begin{split}
\mu_i^{(l)}
= [\sum_{j\in\B(i)}\big[C_{j\arrow i}^{\ast}\big]^{-1}]^{-1}
\Big\{
 \sum_{j\in\B(i)}\big[
C_{j\arrow i}^{\ast}\big]^{-1}(r_{ij}-\mu_j^{(l-1)})\Big\}.
\end{split}
\end{equation}
In the subsequent, we prove the following theorem for the convergence property of $\mu_i^{(l)}$.
\begin{mytheorem}
For asynchronous updating, with feasible  initial $P_j^{(0)}$,
the mean of LSBP algorithm, i.e., $\mu_i^{(l)}$ in (\ref{dsbp}), converges to a
fixed point irrespective of the network topology.
\end{mytheorem}
\begin{myproof}
Let
$K_{ji} \triangleq  { [\sum_{j\in\B(i)}\big[C_{j\arrow i}^{\ast}\big]^{-1}]^{-1}} {\big[
C_{j\arrow i}^{\ast}\big]^{-1} },
$
and
$\xi_i \triangleq[\sum_{j\in\B(i)}\big[ C_{j\arrow i}^{\ast}\big]^{-1}]^{-1}
\Big\{
 \sum_{j\in\B(i)}\big[
C_{j\arrow i}^{\ast}\big]^{-1}r_{ij}\Big\}
$,
then (\ref{dsbp}) can be expressed as
\begin{equation}\label{beliefu2}
\mu_i^{(l)} =\xi_i -
 \sum_{j\in\B(i)}
 K_{j,i} \mu_j^{(l-1)}.
\end{equation}
Due to the fact that $f_1$ is the reference for pre-compensated frequency shift estimation, thus $\mu_1^{(l)}$ is a constant which is denoted by $\mu_1$, and then  only the convergence of $\mu_2^{(l)},\mu_3^{(l)},\ldots,\mu_N^{(l)}$ needs to be investigated.
Hence, we separate $\mu_1$ from $\sum_{j\in\B(i)},
 K_{j,i} \mu_j^{(l-1)}$ in (\ref{beliefu2}), and the result can be expressed as
\begin{equation}\label{beliefu3}
\mu_i^{(l)} =(\xi_i - K_{1,i}\mu_1\mathbbm{1}_{1,i})
 -
 \sum_{j\in\{\B(i)\setminus 1\}}
\mathbbm{1}_{j,i} K_{j,i} \mu_j^{(l-1)},
\end{equation}
where
$\mathbbm{1}_{j,i}$ is an indicator random variable with $\mathbbm{1}_{j,i}=1$ if $\{j,i\}\in \mathcal{E}$ otherwise is $\mathbbm{1}_{j,i}=0$.

Next, the convergence of $\mu_2^{(l)},\mu_3^{(l)},\ldots,\mu_N^{(l)}$ will be investigated all together.
Define $\bm \mu^{(l)} = [\mu_2^{(l)},\mu_3^{(l)},\ldots,\mu_N^{(l)}]^T$,
and $\bm k_i = [\mathbbm{1}_{2,i} K_{2,i},\mathbbm{1}_{2,i} K_{2,i},\ldots, \mathbbm{1}_{N,i} K_{N,i}]^T$, and then
(\ref{beliefu3}) can be reformulated as
\begin{equation}\label{beliefu4}
\mu_i^{(l)} =(\xi_i - K_{1,i}\mu_1\mathbbm{1}_{1,i})
 -
 \bm k_{i}^T \bm\mu^{(l-1)}.
\end{equation}

Piling up (\ref{beliefu4}) for all $\mu_i$ with the increasing order on $i$, we obtain the updating equation for all $ \bm\mu$ as
\begin{equation}\label{beliefu3}
\bm \mu^{(l)} =\bm \eta
 -
 \bm K\bm\mu^{(l-1)} ,
\end{equation}
where $\bm \eta= [\xi_2 - K_{1,2}\mu_1\mathbbm{1}_{1,2},\xi_2 - K_{1,2}\mu_1\mathbbm{1}_{1,2},\ldots]^T$ and $\bm K$ is an $(N-1)\times (N-1)$ matrix with the $i^{\textrm{th}} $ row of $\bm K$ being $\bm k_i^T$.
According to the definition of $\bm k_i$ above (\ref{beliefu4}), the summation of $\bm k_i$ can be written as
$\sum_{j\in \B(i)\setminus 1}K_{j,i}={\sum_{j\in \B(i)\setminus 1}\big[
 \bm C_{j\arrow i}^{\ast}\big]^{-1} }/{ \sum_{j\in\B(i)}\big[\bm C_{j\arrow i}^{\ast}\big]^{-1}}$.
It is obvious that,
if $1\in \B(i)$,
 $\sum_{j\in \B(i)\setminus 1}K_{j,i}< 1$, and
if $1\not\in \B(i)$,
 $\sum_{j\in \B(i)\setminus 1}K_{j,i}\leq 1$.
Therefore, $\bm K$ is a non-negative matrix having row sums less than or equal to $1$ with at least one row sum less than $1$.
Hence, $\bm K$ is a substochastic matrix.
Consequently, $K$   in  (\ref{beliefu3}) is a non-negative and irreducible substochastic matrix, therefore, $\rho(|\bm K|)=\rho(\bm K)<1$, where
 $\rho(\cdot)$ denotes the spectrum radius of a matrix.
Then (\ref{beliefu3}) is convergence guaranteed \cite{MatrixAnalysis}.
Hence, the convergence of $\mu_i^{(l)}$ in (\ref{dsbp}) is guaranteed irrespective the network topology.
\end{myproof}

\section{Experiment Evaluations}\label{simu}
In this section, experiments are conducted to evaluate the proposed algorithm for CFO estimation.
A newwork with $100$ communication agents scatted over a  $3\textrm{km}\times 4\textrm{km}$ area is studied.
Doppler shift is used for the simulation of
the true CFO between each pair of agents within communication range according to  $f_{i}+f_{j}=v_{i,j}f_0/c$, where $v_{i,j}$ is the relative velocity between agents $i$ and $j$, $f_0$ is the carrier frequency, and $c$ is the speed of waves.

In practice, message exchange between agents may fail due to
various factors, like separation distance, signal propagation
environment,
received signal strength, transmission power and modulation
rate \cite{du2017proactive}.
In the following experiments, different packet delivery ratio (PDR), which is the ratio of the number of packets successfully delivered to destination compared to the number of packets that have been sent out by the transmitter, is set to show the impacts of packet drop on proposed algorithms.

\begin{figure}[t]
  \centering
{\epsfig{figure=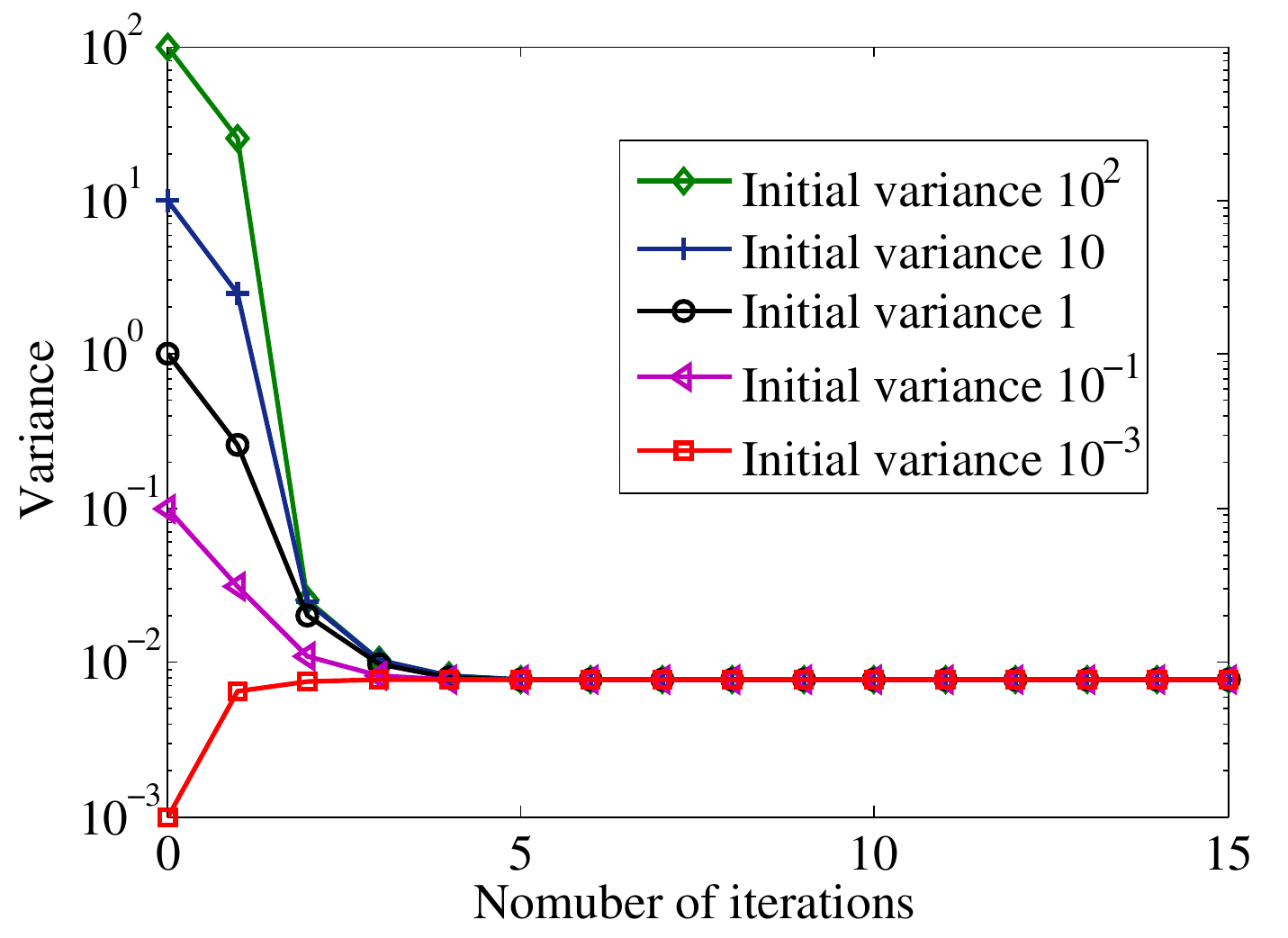,width=2.7in}}
\caption{Convergence property of $\bm P_6^{(l)}$  for different initial values.  }
\label{1234}
\end{figure}

First, the convergence property of $ P_{i}^{(l)}$ as shown in Theorem \ref{ConConvTh} is verified by simulations.
The network topology is randomly generated, and PDR is set to be $80\%$.
The initial message variance for each
$ P_{i}^{(0)}$  is set to be $100$, $10$, $1$, $0.1$ and $0.01$, respectively.
The convergence property of $ P_{6}^{(l)}$ is demonstrated in Fig. \ref{1234} as an example.
It is clear that though $ P_{6}^{(l)}$ keeps monotonic increasing or decreasing with different initial values, they converge to the same point fast.
Thus, the conclusion of Theorem \ref{ConConvTh} is verified by simulations that
with arbitrary feasible initial value, the belief variance $ P_i^{(l)}$ of LSBP shown in (\ref{beliefP2})  converges to a unique fixed point.
And from Theorem \ref{ConConvTh}, we know the convergence rate is doubly exponential.

\begin{figure}[t]
  \centering
{\epsfig{figure=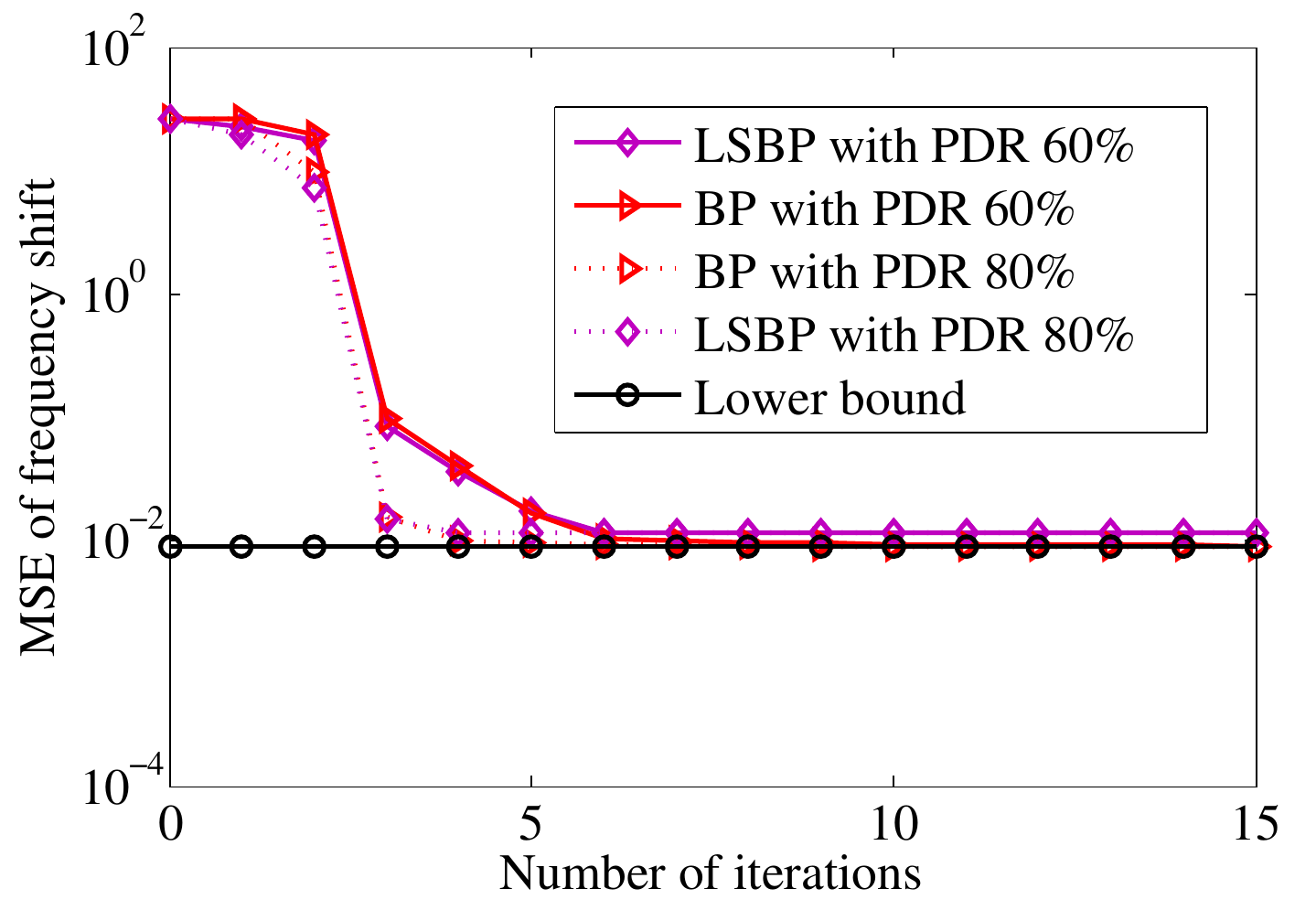,width=2.7in}}
\caption{Accuracy and convergence property of $\hat{f}_i$ under different PDRs.}
\label{MSE-Iter-PDR}
\end{figure}

Next, the accuracy and convergence property of $\hat{f}_i$ is investigated.
Average MSE, defined as
$\frac{1}{N}\sum_{i=1}^{N}\mathbb{E}\{(\frac{\hat{f}_i-f_i}{B})^2\}$,
is adopted as the performance criteria.
Fig. \ref{MSE-Iter-PDR} shows that for different PDRs ($60\%$ and $80\%$), the convergence speeds of BP and LSBP algorithms differ.
Nevertheless, even for PDR as low as $60\%$, both BP and LSBP converge to a fixed estimate point within $10$ iterations, and thus, they are robust to packet drops.
Besides, LSBP has the MSE performance that approaches the CRLB.
Note that BP can also reach the CRLB as shown in Fig. \ref{MSE-Iter-PDR}, but its convergence for loopy topology network is not guaranteed, and its communication overhead is large.

\begin{figure}[t]
  \centering
{\epsfig{figure=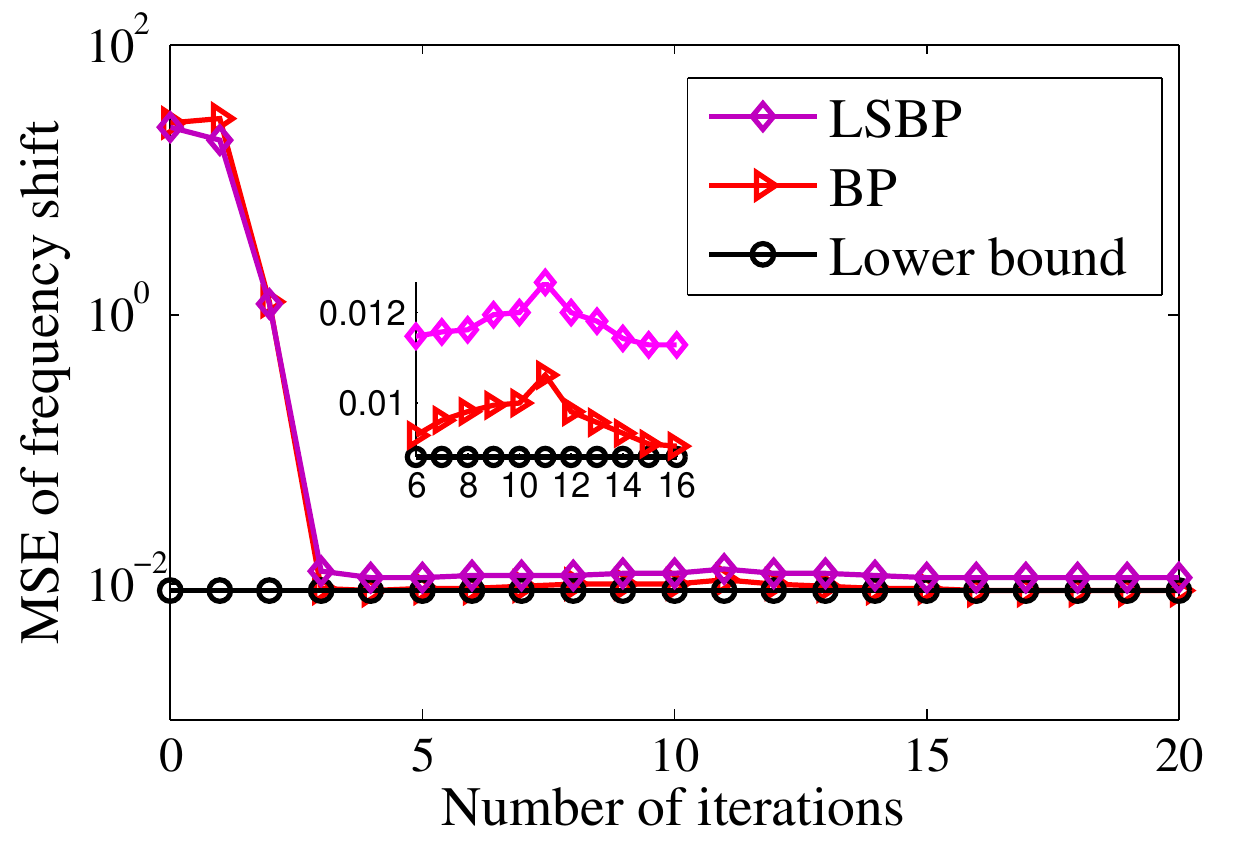,width=2.7in}}
\caption{Adaptive property of proposed algorithms to dynamic  network topology.
At iteration $5$, agent $4$, $5$, $8$ and $10$  leave the network, and at iterations $10$ and $11$,  new agents join the network at former positions of $4$, $5$, $8$ and $10$, respectively.}
\label{MSE-dynamic}
\end{figure}

Fig. \ref{MSE-dynamic} shows adaptiveness property of the proposed algorithms to the dynamic topology of networks.
At first, the network topology is the same as that adopted in Fig. \ref{MSE-Iter-PDR}.
At iteration $5$, agents $4$, $5$, $8$ and $10$  leave the network, and at iterations $10$ and $11$,  new agents join the network at former positions of $4$, $5$, $8$ and $10$, respectively.
It can be seen that the average MSE increases at iteration $6$ due to agents' leaving, and it decreases after iteration $11$ because new agents join in and bring new measurements.
It is shown that the impact of agents' leaving and joining on the performance of BP  and LSBP  is very trivial, and both algorithms are adaptive to topology varying.

\section{Conclusions}\label{conclusion}
We have studied a distributed message passing algorithm, named as linear scaling belief propagation (LSBP), for distributed frequency synchronization, where the communication overhead is linear scaling with the network density.
Analytical analysis has been conducted to rigorously prove that the the proposed algorithm is convergence guaranteed with feasible initial values even for systems with packet drops and random delays.
Though LSBP only requires local information at each agent, simulationshave verified that LSBP achieves almost the optimal frequency compensation accuracy with an error approaching the Cram\'{e}r-Rao lower bound.
Simulations also show that the number of exchanged messages linearly scales with the number of agents, and the iteration number upon convergence increases mildly, and thus, implementing LSBP imposes tolerable communication overhead.


\end{document}